\documentclass[floatfix,prl,twocolumn,showpacs,amsmath,amssymb,letterpaper,groupaddresses,superscriptaddress]{revtex4}
\usepackage{times}
\usepackage{latexsym}
\usepackage{graphicx}
\usepackage{verbatim,times,bbm}
\usepackage{color}

\usepackage{color}

\begin{document}

\title{Enhanced Quantum Communication via Optical Refocusing}

\author{Cosmo Lupo}
\affiliation{School of Science and Technology, University of Camerino, via Madonna delle Carceri 9, I-62032 Camerino, Italy}

\author{Vittorio Giovannetti}
\affiliation{NEST, Scuola Normale Superiore and Istituto Nanoscienze-CNR, piazza dei Cavalieri 7, I-56126 Pisa, Italy}

\author{Stefano Pirandola}
\affiliation{Department of Computer Science, University of York, York YO10 5GH, UK}

\author{Stefano Mancini}
\affiliation{School of Science and Technology, University of Camerino, via Madonna delle Carceri 9, I-62032 Camerino, Italy}
\affiliation{INFN-Sezione di Perugia, I-06123 Perugia, Italy}

\author{Seth Lloyd}
\affiliation{Department of Mechanical Engineering, MIT Cambridge, MA 02139, USA}

\begin{abstract}

We consider the problem of quantum communication mediated by an 
optical refocusing system, which is schematized as a thin lens 
with a finite pupil.
This model captures the basic features of all those situations 
in which a signal is either refocused by a repeater for long 
distance communication, or it is focused on a detector prior to 
the information decoding process. 
Introducing a general method for linear optical systems, we compute 
the communication capacity of the refocusing apparatus.
Although the finite extension of the pupil may cause loss of 
information, we show that the presence of the refocusing system 
can substantially enhance the rate of reliable communication 
with respect to the free-space propagation.

\end{abstract}


\maketitle

Although quantum information is more commonly described in terms of 
discrete variables (e.g., qubits), information is most naturally encoded 
in the electromagnetic field (EMF) via a continuous variable representation~\cite{CVs}.
All the fundamental quantum information tools and protocols, 
from teleportation to quantum key distribution, have been demonstrated 
for such encoding~\cite{tools}.
Here we consider the problem of quantum 
communication~\cite{QOchannel,Shapiro,YUENSH} and compute 
the maximum rate at which information
can be reliably transmitted through EMF signals which 
propagate along an optical communication line under refocusing conditions.
Even though we explicitly consider classical information~\cite{Ccapacity}, 
our results are immediately extensible to the case of quantum 
information~\cite{Qcapacity}.

In the classical domain, the ultimate limits for communication via 
continuous variable encodings were provided by the seminal work of 
C.~E.~Shannon on Gaussian channels~\cite{Shannon}. 
In the quantum domain such channels are replaced by the so called 
Bosonic Gaussian Channels (BGC), which describe the propagation of the 
EMF in linear media~\cite{HOLWER}.
Their structure is notably rich~\cite{Gchannels}, and up to date a full 
information-theoretical characterization has been achieved only for certain 
special subclasses~\cite{broadband,Wolf}.
These results have been applied to compute the maximal rates of reliable 
communication via attenuating media, as optical fibers, wave-guides, and 
via free-space propagation~\cite{freespace,Shapiro,wave}. 

Here we move further in this direction by characterizing the propagation
of the EMF through a linear optical system,
which induces refocusing of the transmitted signals.
For the sake of simplicity, we model this apparatus as a thin lens with
finite pupil placed between the sender of the message and the 
receiver under focusing conditions. 
Notwithstanding its relatively simple structure, this model captures the 
basic features of all those situations in which a signal is either 
refocused by a suitable repeater to allow long distance communication, 
e.g., by means of parabolic antenna for satellite 
communication~\cite{satellite}, or it is focused on a detector 
prior to the information decoding process. 
The latter include, e.g., the settings in which the EMF is used for the 
readout process and those in which an optical system is used to 
interface light and matter~\cite{lenses}. 

The quantum description of the scattering of the signal by the optical 
system is derived by consistently applying the canonical quantization rules, 
see~\cite{Shapiro} and Refs.\ therein.
Within this framework we show that while the finiteness of the pupil
limits the channel bandwidth~\cite{telegraph}, the rate of information 
transmission is always enhanced by the presence of the lens. 
It turns out that the resulting improvement with respect to the free-space 
communication scheme may be particularly advantageous when 
only few photons per mode are employed in the information transmission --- 
a configuration which is close to the operative regimes of the 
long distance free-space communication protocols realized so far~\cite{longd}.

{\it The optical system.--} Consider a linear optical system 
with a set of transmitter modes, labelled by $i$, and receiver 
modes, labelled by $j$.  
In the case of RF or microwave communication, for example,
the transmitter and receiver could be antennae.  For optical   
communication, the transmitter could be a laser coupled to 
a telescope, and the receiver could be a telescope coupled
to a CCD array.  Transmitter and receiver modes typically
have both spatial characteristics determined by the optical
characteristics of the transmitter and receiver, and temporal
characteristics, determined by the frequency and bandwidth
of the transmitted radiation.  
The transmittivity matrix $T_{ji}$ gives the
fraction of light from the $i$'th transmitter mode that is
received at the $j$'th receiver mode.  We would like to
determine the maximum amount of information that can
be sent from transmitter to receiver for fixed total
input power.

Let us consider the purely lossy case, in which noise from
the environment is negligible. This is the case, for example,
for free-space optical communication in a thermal background.
The addition of noise will be considered below.
If the loss is $\eta$ and there are $\nu$ parallel channels 
with total average photon number $N$, then the communication
capacity, measured in nats, is~\cite{broadband}
\begin{equation}\label{capacity}
C = \nu g(\eta N/\nu) \, ,
\end{equation} 
where $g(x):=(x+1)\ln{(x+1)}-x\ln{x}$, and the capacity is 
attained by sending coherent states down the channel. 

In our case, we have a single multimode lossy channel with
transfer matrix $T_{ji}$, which mixes the input modes together. 
This channel can be transformed into a set of parallel channels 
by using the singular value decomposition. 
The singular value decomposition states that any matrix $T$ 
can be written as $T = {\cal V} \Sigma {\cal U}$, where 
${\cal V}$, ${\cal U}$ are unitary matrices, and $\Sigma$ 
is a nonnegative diagonal matrix. 
We can write $T$ in components as
$T_{ji} = \sum_k {\cal V}_{jk} \sqrt{\eta_k} {\cal U}_{ki}$.
The $\sqrt{\eta_k}$ are the singular values of the transfer
matrix. The singular value decomposition shows that any multimode
lossy channel can be decomposed into parallel lossy channels
with input modes corresponding to the rows of ${\cal U}$, 
output modes corresponding to the columns of ${\cal V}$, 
and loss factors corresponding to the singular values $\eta_k$.
The singular value input modes can now be quantized using 
annihilation and creation operators $a_i, a^\dagger_i$: 
$[a_i, a^\dagger_{i'}] = \delta_{i,i'}$.
Similarly, the output modes can be quantized using 
operators $b_i, b^\dagger_i$: $[b_i, b^\dagger_{i'}] = \delta_{i,i'}$.
To preserve the canonical quantization relationships, each
input-output pair is coupled to a loss mode with operators 
$\xi_i, \xi^\dagger_i$, $b_i = \sqrt{\eta_i} a_i + \sqrt{1-\eta_i} \xi_i$.
We see that the singular value decomposition of the multimode
lossy quantum channel renders the channel completely equivalent
to a set of parallel lossy quantum channels with loss factors
$\eta_i$. 

\begin{figure}
\centering
\includegraphics[width=0.4\textwidth]{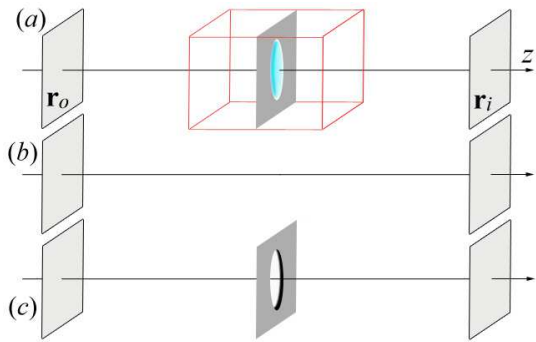}
\caption{(color online) ($a$) Scheme for optical communication
through an optical refocusing apparatus, modeled as a thin lens
of radius $R$ and focal length $f$.
($b$) Free-space propagation scenario.
($c$) Alternative scenario in which the lens is replaced by a
hole of the same size in the absorbing screen.
$\mathbf{r}_o$ and $\mathbf{r}_i$ denote the Cartesian
coordinates on the object and image planes.} \label{optics}
\end{figure}

We consider the case of monochromatic light propagating along an optical axis. 
Following~\cite{Shapiro,YUENSH} the input and output signals are identified 
by the transverse field modes at two planes orthogonal to the optical axis 
[the object plane and image plane of Fig.~\ref{optics}($a$)], and the field 
propagation is defined by assigning the point-spread function (PSF) 
$T(\mathbf{r}_i,\mathbf{r}_o)$ which connects the field amplitude at position 
$\mathbf{r}_o$ on the first plane with the field amplitude at position $\mathbf{r}_i$ 
on the second one~\cite{Goodman}. 
Due to diffraction, such inputs are scattered over the whole image 
plane according to amplitude probability distributions defined by the PSF. 
This setting formally defines a Gaussian memory channel~\cite{unravel}, in which 
output signals originated by distinct input fields are not mutually independent.

As mentioned in the introduction, we model the optical refocusing system as a 
converging lens of focal length $f$, located at distance $D_o$ from 
the object plane. 
Working in the thin-lens approximation, and neglecting aberrations, 
light is focused at the image plane located at distance $D_i$ from the 
optical system, where $1/D_o + 1/D_i = 1/f$. 
Eventually the image is magnified by a factor $M = D_i/D_o$. 
Diffraction of light is responsible for image blurring and causes loss 
of information. It can be described by introducing an effective entrance 
pupil characterizing the optical system. 
Denoting $P(\mathbf{r})$ the characteristic function of the pupil that 
encircles the lens, the PSF for the monochromatic field at 
wavelength $\lambda$ is obtained, in the paraxial approximation, 
by Fourier transforming $P(\mathbf{r})$~\cite{Goodman}. 
For a circular pupil of radius $R$, the PSF reads
\begin{equation}\label{transfer}
T(\mathbf{r}_i,\mathbf{r}_o) =
\frac{e^{j\vartheta(\mathbf{r}_i,\mathbf{r}_o)} R^2}{\lambda^2 D_o
D_i} \frac{J_1(2\pi R \rho )}{R\rho} \, ,
\end{equation}
where $J_1$ is the Bessel function of first kind and order one,
$\vartheta(\mathbf{r}_i,\mathbf{r}_o) = \frac{\pi}{\lambda D_o}
\left( |\mathbf{r}_o|^2 + |\mathbf{r}_i|^2/M \right) + \frac{2\pi
D_o}{\lambda}\left( 1+M \right)$, and $\rho = |\mathbf{r}_i-M\mathbf{r}_o|/(\lambda D_i)$.
The PSF in~(\ref{transfer}) accounts for two physical phenomena:
(i)~The focusing properties of the converging lens; 
(ii)~The loss of the field components impinging outside the pupil of the 
optical system.
In other words, Eq.~(\ref{transfer}) assumes the presence of an absorbing
screen surrounding the lens.
Eventually, one could derive a PSF describing the propagation 
through a lens of a radius $R$ which is not surrounded by an absorbing 
screen, allowing the transfer of the field components which are not 
refocused by the lens. 
However, for the sake of conciseness, in the following we adopt the 
expression~(\ref{transfer}) and show that the presence of the converging 
lens increases the communication capacity, with respect to the free-space 
propagation, in the settings in which the signal loss caused by the 
absorbing screen is negligible.

To characterize our optical refocusing system, we apply the singular
value decomposition to the PSF, which plays the role of the transfer
matrix.
The system is hence characterized by a set of loss factors.
In the farfield and nearfield limits, they can be computed exactly 
in terms of the Fresnel number associated to the optical system \cite{details}.
To fix the ideas, let us assume that information is encoded in the object 
plane on a square of length $L$, creating an image on the image plane 
which (in the geometric optics approximation) is roughly contained in a 
square of size $ML$. 
The Fresnel number associated to this setup is
\begin{equation}
\mathcal{F} = \frac{\pi R^2 L^2}{\lambda^2 D_o^2} = \pi\left(\frac{L}{x_\mathsf{R}}\right)^2 \, ,
\end{equation}
$x_\mathsf{R}= \lambda D_o/R$ being the Rayleigh length of the system. 
In the farfield limit, $\mathcal{F} \ll 1$, only one mode is transmitted
with loss $\eta = \mathcal{F}^2$. 
In the nearfield limit, $\mathcal{F} \gg 1$, a number $\nu=\mathcal{F}$ of
modes are transmitted without loss. 

To evaluate the effects of the refocusing system (that is, the converging 
lens) on the information transmission, we use the free-space propagation
of the EMF as a term of comparison, see Fig.~\ref{optics}($b$). 
The characterization of the free-space propagation of the field, including
the quantum regime, can be found in~\cite{Shapiro} and Refs.\ therein.
For this scenario, the associated Fresnel number is 
$\mathcal{F}_\mathrm{fs} = (A_1 A_2)/(\lambda d)^2$,
where $A_1$ and $A_2$ are the areas of the surface from which the signal 
is emitted and on which it is detected, and $d$ is the distance 
between them.
For a fair comparison, we use for both the scenarios the same values for 
areas of the input and output surfaces and their distance, yielding the 
free-space Fresnel number 
\begin{equation}
\mathcal{F}_\mathrm{fs} = \frac{M^2L^4}{\lambda^2 D^2} = \frac{L^4}{\lambda^2 D_o^2}\left(\frac{M}{1+M}\right)^2 \, ,
\end{equation}
where $D=D_o+D_i=D_o(1+M)$.
Also in this case, it is possible to derive exact expressions for the 
effective transmissivities in the farfield and nearfield limit:
In the farfield limit, $\mathcal{F}_\mathrm{fs} \ll 1$, only one mode is 
transmitted, with loss $\eta_\mathrm{fs}=\mathcal{F}_\mathrm{fs}$;
In the nearfield limit, $\mathcal{F}_\mathrm{fs} \gg 1$, a number of modes 
$\nu_\mathrm{fs}=\mathcal{F}_\mathrm{fs}$ are transmitted without losses.


{\it Optical refocusing vs free-space propagation.--} First of all we
notice that, for given values of the physical parameters, the two
scenarios, in the following denoted by ($a$) and ($b$), can independently 
operate in the farfield or nearfield limit, or in none of the two.

Let us consider first the case in which both the scenarios operate in the 
farfield regime. The ratio between the loss factors of the transmitted
modes equals
\begin{eqnarray}
r_1 =  \frac{\eta}{\eta_\mathrm{fs}} 
= \left(\frac{\pi R^2}{\lambda D_o}\right)^2 \; \left( \frac{1+M}{M} \right)^2\;,  
\end{eqnarray}
which is larger than one only if the parameters $M, R, D_o, \lambda$ do 
satisfy a certain condition. 
We now show that such a condition is fulfilled when the loss of signal 
induced by absorbing screen surrounding the pupil is negligible.
To do so, we consider a third scenario, denoted by ($c$) in Fig.~\ref{optics}, 
in which the field propagates from the object plane to the image plane 
and the converging lens is replaced by a hole of the same size in the 
absorbing screen.
The field propagation in this configuration can be analyzed by splitting 
it in two parts: the free-space propagation from the object plane to the 
screen ($o \rightarrow s$) and the one from the screen to the image plane 
($s \rightarrow i$).
These two free-space propagations are associated to the Fresnel numbers
$\mathcal{F}_\mathrm{fs}^{o \rightarrow s} = \pi R^2 L^2/\lambda^2 D_o^2 =\mathcal{F}$,
and $\mathcal{F}_\mathrm{fs}^{s \rightarrow i} = \pi R^2 (ML)^2/\lambda^2 D_i^2 = \mathcal{F}$.
(Notice that they are identical due to the fact that $ML/D_i= L/D_o$.)
Hence the farfield condition on the scenario ($a$) implies that both 
those propagations take place in the farfield regime as well.
It follows that there is, in the scenario ($c$), at most one mode 
propagating $o \rightarrow s \rightarrow i$, which is attenuated by a 
factor 
\begin{eqnarray}
\eta^{o \rightarrow s \rightarrow i} \leqslant \mathcal{F}_\mathrm{fs}^{o \rightarrow s} \mathcal{F}_\mathrm{fs}^{s \rightarrow i} = \mathcal{F}^2 = \eta \; .
\end{eqnarray} 
Now, if the presence of the absorbing screen around the pupil is negligible, 
we must have that the losses on the $o \rightarrow s \rightarrow i$ 
propagation (quantified by the factor $\eta^{o \rightarrow s \rightarrow i}$)
should be equal to those gotten from direct free-space propagation
(quantified by $\eta_\mathrm{fs}$). 
But since $\eta^{o \rightarrow s \rightarrow i}$ is not greater than 
$\eta$, it follows that the regime in which we can neglect the effect of 
the absorbing screen is the one in which $r_1 \geqslant 1$. 
In other words, the detrimental effects that we see for $r_1 < 1$ merely 
correspond to the absorptions by the screen. 
As a final remark we also observe that the condition $r_1 >1$, plus the 
farfield condition for the scenario ($a$), enforces the farfield regime 
for the scenario ($b$).
Finally, we compare the performances of the two scenarios in terms 
of capacity by computing the gain
\begin{eqnarray}
G_1 = \frac{C}{C_\mathrm{fs}} = \frac{g(\eta N)}{g(\eta_\mathrm{fs} N)} = \frac{g(r_1 \eta_\mathrm{fs} N)}{g(\eta_\mathrm{fs} N)} \; .
\end{eqnarray}
We notice that in the semiclassical limit, $N \gg 1$, the gain satisfies 
$G_1 \simeq 1$, that is, the presence of the optical refocusing system 
does not affect the information transmission capacity~\cite{glimit}. 
On the other hand, the gain can be significantly greater than $1$ 
in the quantum regime: 
In particular, the gain is maximum in the limit of weak signals, $N \ll 1$, 
in which $G_1 \simeq r_1$.
 
Let us now move to the case in which both the scenarios operate in the 
nearfield regime.
The ratio between the number of modes perfectly transmitted in the two 
scenarios is
\begin{equation}
r_2 = \frac{\nu}{\nu_\mathrm{fs}} = \pi \left(\frac{R}{L}\right)^2 \; \left( \frac{1+M}{M} \right)^2 \; ,
\end{equation}
which can be larger or smaller than one, depending on the geometric 
parameters $M$, $R$, $L$.
However, as in the previous case, we show that if the losses induced by 
the absorbing screen are negligible, then $r_2 \geqslant 1$.
Again, to show that let us consider what happens in the scenario ($c$). 
We first notice that the nearfield condition for the scenario ($a$)
implies that both $o \rightarrow s$ and $s \rightarrow i$ 
propagations are in the nearfield regime.
The number of transmitted modes are 
$\nu_\mathrm{fs}^{o \rightarrow s} = \pi L^2 R^2/\lambda^2 D_o^2= \nu$
and $\nu_\mathrm{fs}^{s \rightarrow i} = \pi (ML)^2 R^2/\lambda^2 D_i^2= \nu$.
Hence the number of modes that propagate from the object to the image plane 
in the scenario ($c$) satisfies the inequality
\begin{equation}\label{nuosi}
\nu^{o \rightarrow s \rightarrow i} \leqslant \min\{\nu_\mathrm{fs}^{o \rightarrow s}, \nu_\mathrm{fs}^{s \rightarrow i}\} = \nu \;.
\end{equation} 
It is clear that the presence of the pupil is negligible only if the 
number of modes transmitted in the propagation $o \rightarrow s \rightarrow i$ 
equals the number of those transmitted in the free-space propagation, 
that is, if $\nu^{o \rightarrow s \rightarrow i} \simeq \nu_\mathrm{fs}$.
Equation~(\ref{nuosi}) implies that this is the setting for which 
$r_2 \geqslant 1$.
The ratio between the classical capacities of the corresponding quantum 
channels reads 
\begin{equation}
G_2 = \frac{C}{C_\mathrm{fs}} = \frac{\nu \; g(N/\nu)}{\nu_\mathrm{fs} \;g(N/\nu_\mathrm{fs})} = r_2  \frac{g(N/\nu)}{g(r_2 N/\nu)} \; .
\end{equation}
Notice that the ratio $N/\nu$ represents the number of photons per 
transmitted mode.
In the limit $N/\nu \gg 1$ we are in the semiclassical regime, for 
which the gain is $G_2 \simeq r_2$.
In the quantum regime $N/\nu \simeq 1$ we have $G_2 > 1$.
Finally, in the limit of weak signals, $N/\nu \ll 1$, the gain tends to 
$G_2 \simeq 1$.

One may notice that the condition $r_2 \geqslant 1$, together with the 
nearfield condition for the scenario ($a$), is not sufficient to infer 
the nearfield condition for the scenario ($b$). 
Hence we shall compare the nearfield case for the scenario ($a$) with 
the farfield case for the scenario ($b$), a setting which is 
characterized by the condition
$\frac{L^2}{D_o} \frac{M}{M+1} \ll \lambda \ll \frac{LR}{D_o}$.
In this case the gain becomes 
\begin{equation}
G_3 = \frac{C}{C_\mathrm{fs}} = \frac{\nu \; g(N/\nu)}{ \;g(\eta_\mathrm{fs} N)} \; ,
\end{equation}
which, in the semiclassical limit, $N \gg 1$, is $G_3 \simeq \nu \gg 1$, 
and for very weak signals, $N \ll 1$, is $G_3 \simeq 1/\eta_\mathrm{fs} \gg 1$.

The enhancement in the transmission rate provided by the optical refocusing 
system persists in the presence of thermal noise.
In such a case, by encoding classical information into coherent states,
Eq.~(\ref{capacity}) has to be replaced by
$C = g(\eta N/\nu + N_\mathrm{th})-g(N_\mathrm{th})$~\cite{thermal,HOLWER}, 
where $N_\mathrm{th}$ is the number of thermal photons per transmitted mode,
which yields $G_1 \simeq r_1$, $G_3 \simeq 1/\eta_\mathrm{fs}$ for
$N_\mathrm{th} \gg \max\{1, \eta N/\nu\}$, and $G_2 \simeq r_2$, $G_3 \simeq \nu$
for $1 \gg N_\mathrm{th} \gg \eta N/\nu$.

{\it Conclusions.--} We have computed the capacity of quantum optical 
communication through an optical refocusing system, modeled as a thin
lens with finite pupil.
Despite its simplicity, the model is general enough to find application
in different contexts, from refocusing antennas for long-distance 
communication to imaging systems on the small and medium scale, and 
accounts for the focusing process, light diffraction and power loss.
We have shown that, under certain conditions, the converging optical 
apparatus can be used to achieve, in comparison with the free-space 
field propagation, higher transmission rates.
Our results furnish the ultimate limits of quantum optical communication 
and may be useful for determining general bounds on the efficiency 
of any protocol requiring the transmission of quantum degrees of freedom 
of light, e.g., quantum imaging~\cite{GLMS} and quantum discrimination~\cite{discrimine}. 

\acknowledgments 
The research leading to these results has received
funding from the European Commission's seventh Framework Programme
(FP7/2007-2013) under grant agreements no.~213681, and by the
Italian Ministry of University and Research under the FIRB-IDEAS
project RBID08B3FM.  
V.G. also acknowledges the support of Institut Mittag-Leffler (Stockholm), 
where he was visiting while part of this work was done.
The authors thank Lorenzo Maccone for useful discussions and comments,
C.L. warmly thanks Ciro Biancofiore for his valuable scientific support.


\begin{thebibliography}{99}
\bibitem{details}
Details of the calculation will be presented elsewhere.
\bibitem{glimit}
To show this use the fact that $g(x) \simeq \ln(x)+1$ for $x\gg1$.
\end{thebibliography}
\end{document}